\documentclass[prd,showpacs,amsmath,amssymb,floatfix]{revtex4}
\usepackage{graphicx,color,dcolumn,booktabs}

\begin{document}
\title{Possible heavy molecular states composed of a pair of excited charm-strange mesons}

\author{Bin Hu}
\author{Xiao-Lin Chen}
\email{chenxl@pku.edu.cn}
\author{Zhi-Gang Luo}
\affiliation{Department of Physics
and State Key Laboratory of Nuclear Physics and Technology\\
and Center of High Energy Physics, Peking University, Beijing
100871, China }
\author{Peng-Fei Yu}\author{Xiang Liu}\email{xiangliu@lzu.edu.cn}
 \affiliation{School of Physical
Science and Technology, Lanzhou University, Lanzhou 730000, China}

\author{Peng-Zhi Huang}
\author{Shi-Lin Zhu}\email{zhusl@pku.edu.cn}
\affiliation{Department of Physics
and State Key Laboratory of Nuclear Physics and Technology\\
and Center of High Energy Physics, Peking University, Beijing
100871, China }

\date{\today}

\begin{abstract}

The P-wave charm-strange mesons $D_{s0}(2317)$ and $D_{s1}(2460)$
lie below the $DK$ and $D^\ast K$ threshold respectively. They are
extremely narrow because their strong decays violate the isospin
symmetry. We study the possible heavy molecular states composed of
a pair of excited charm strange mesons. As a byproduct, we also
present the numerical results for the bottonium-like analogue.

\end{abstract}


\maketitle

\section{Introduction}\label{sec1}

In the past seven years, the experimentally observed new
charmonium or charmonium-like states include $X(3872)$, $Y(3940)$,
$Y(4260)$, $Z(3930)$, $X(3940)$, $Y(4325)$, $Y(4360)$, $Y(4660)$, $Z^+(4430)$,
$Z^+(4050), Z^+(4250)$ and $Y(4140)$ etc
\cite{2003-Choi-p262001-262001,2005-Choi-p182002-182002,
2005-Aubert-p142001-142001, 2006-Uehara-p82003-82003,
2007-Abe-p82001-82001, 2007-Aubert-p212001-212001,
2007-Yuan-p182004-182004, 2007-Wang-p142002-142002,
2008-Choi-p142001-142001, 2008-Mizuk-p72004-72004,
2009-Aaltonen-p242002-242002}. It's difficult to accommodate all
these states especially those charged ones in the conventional
quark model. Many of these new states lie close to the threshold
of two charmed mesons. A natural speculation is that some of them
may be the molecular states composed of two charmed mesons
\cite{2004-Swanson-p197-202,
2008-Liu-p94015-94015,2009-Liu-p17502-17502,liu1,liu2,liu3,liu4,liu5}.

In the framework of the meson exchange model, we have investigated
the possible loosely bound molecular states composed of a pair of
the ground state S-wave heavy mesons and a pair of S-wave and
P-wave heavy mesons in Refs. \cite{liu1,liu2,liu3,liu4,liu5}. In
this work we go one step further and study the possible molecular
system composed of a pair of P-wave heavy mesons in the $(0^+,
1^+)$ doublet according to the classification of the heavy quark
symmetry.

The non-strange P-wave $(0^+, 1^+)$ heavy mesons are very broad
with a width around several hundred MeV \cite{pdg}. Instead of
forming a stable molecular state, the system composed of a pair of
non-strange P-wave heavy meson decays rapidly. Experimental
identification of such a molecular state will be very difficult.
The attractive interaction between the meson pairs may lead to a
possible threshold enhancement in the production cross section.

In contrast, the P-wave charm-strange mesons $D_{s0}(2317)$ and
$D_{s1}(2460)$ lie below the $DK$ and $D^\ast K$ threshold
respectively. They are extremely narrow because their strong
decays violate the isospin symmetry. The future experimental
observation of the possible heavy molecular states composed of a
pair of excited charm strange mesons may be feasible if they
really exist. We study the charmonium-like system composed of a
pair of excited charm strange mesons in this work.

This paper is organized as follows. We review the formalism in
Section \ref{sec2} and present the results in Section
\ref{sec3}-\ref{sec5}. The last section is a short summary.

\section{Formalism}\label{sec2}

\subsection{Flavor wave functions}

We list the flavor wave functions of the possible molecular states
composed of the P-wave $(0^+, 1^+)$ heavy doublet in Table
\ref{sec2-tab1}-\ref{sec2-tab2}. $D^*_0$ denotes
$(D^{*0}_0,D^{*+}_0,D^{*+}_{s0})$ while $D_1$ denotes
$(D^0_1,D^+_1,D^+_{s0})$. The neutral $D^*_0-\bar{D}_1$ system
with the parameter $c =\pm 1$ corresponds to the positive and
negative charge parity respectively.

\begin{table}[htb]
\centering \caption{The flavor wave function of the
$D^*_0-\bar{D}^*_0$ and $D_1-\bar{D}_1$ system.}\label{sec2-tab1}
\begin{tabular}{cc|cc}
\hline\hline
& $D^*_0-\bar{D}^*_0$ & & $D_1-\bar{D}_1$ \\
\hline
state & wave function & state & wave function\\
\hline
$\Phi^+$ & $D^{*+}_0\bar{D}^{*0}_0$ & $\Phi^{**+}$ & $D^+_1\bar{D}^0_1$ \\
$\Phi^-$ & $D^{*-}_0D^{*0}_0$ & $\Phi^{**-}$ & $D^-_1D^0_1$ \\
$\Phi^0$ &
$\frac{1}{\sqrt{2}}(D^{*0}_0\bar{D}^{*0}_0-D^{*+}_0D^{*-}_0)$ &
$\Phi^{**0}$ & $\frac{1}{\sqrt{2}}(D^0_1\bar{D}^0_1-D^+_1D^-_1)$ \\
$\Phi^0_8$ & $\frac{1}{\sqrt{2}}(D^{*0}_0\bar{D}^{*0}_0+D^{*+}_0D^{*-}_0)$
& $\Phi^{**0}_8$ & $\frac{1}{\sqrt{2}}(D^0_1\bar{D}^0_1+D^+_1D^-_1)$ \\
$\Phi^+_s$ & $D^+_{s0}\bar{D}^*_0$ & $\Phi^{**+}_s$ & $D^+_{s1}\bar{D}^0_1$ \\
$\Phi^-_s$ & $D^{*-}_{s0}D^{*0}_0$ & $\Phi^{**-}_s$ & $D^-_{s1}D^0_1$ \\
$\Phi^0_s$ & $D^{*+}_{s0}D^{*-}_0$ & $\Phi^{**0}_s$ & $D^+_{s1}D^-_1$ \\
$\bar{\Phi}^0_s$ & $D^{*-}_{s0}D^{*+}_0$ & $\bar{\Phi}^{**0}_s$ & $D^-_{s1}D^+_1$ \\
$\Phi^0_{s1}$ & $D^{*+}_{s0}D^{*-}_{s0}$ & $\Phi^{**0}_{s1}$ & $D^+_{s1}D^-_{s1}$ \\
\hline\hline
\end{tabular}
\end{table}

\begin{table}[htb]
\centering \caption{The flavor wave function of the
$D^*_0-\bar{D}_1$ system. The parameter $c =\pm 1$ for the
$D^*_0-\bar{D}_1$ system with positive and negative charge parity
respectively.}\label{sec2-tab2}
\begin{tabular}{c|c}
\hline\hline
& $D^*_0-\bar{D}_1$ \\
\hline
state & wave function \\
\hline
$\Phi^{*+}/\widehat{\Phi}^{*+}$ & $\frac{1}{\sqrt{2}}(D^{*+}_0\bar{D}^0_1+cD^+_1\bar{D}^{*0}_0)$ \\
$\Phi^{*-}/\widehat{\Phi}^{*-}$ & $\frac{1}{\sqrt{2}}(D^{*0}_0D^-_1+cD^0_1D^{*-}_0)$ \\
$\Phi^{*0}/\widehat{\Phi}^{*0}$ & $\frac{1}{2}[(D^{*0}_0\bar{D}^0_1+cD^0_1\bar{D}^{*0}_0)-(D^{*+}_0D^-_1+cD^+_1D^{*-}_0)]$ \\
$\Phi^{*0}_8/\widehat{\Phi}^{*0}_8$ & $\frac{1}{2}[(D^{*0}_0\bar{D}^0_1+cD^0_1\bar{D}^{*0}_0)+(D^{*+}_0D^-_1+cD^+_1D^{*-}_0)]$ \\
$\Phi^{*+}_s/\widehat{\Phi}^{*+}_s$ & $\frac{1}{\sqrt{2}}(D^{*+}_{s0}\bar{D}^0_1+cD^+_{s1}\bar{D}^{*0}_0)$ \\
$\Phi^{*-}_s/\widehat{\Phi}^{*-}_s$ & $\frac{1}{\sqrt{2}}(D^{*0}_0D^-_{s1}+cD^0_1D^{*-}_{s0})$ \\
$\Phi^{*0}_s/\widehat{\Phi}^{*0}_s$ & $\frac{1}{\sqrt{2}}(D^{*+}_{s0}D^-_1+cD^+_{s1}D^{*-}_0)$ \\
$\bar{\Phi}^{*0}_s/\widehat{\bar\Phi}{}^{*0}_s$ & $\frac{1}{\sqrt{2}}(D^{*+}_0D^-_{s1}+cD^+_1D^{*-}_{s0})$ \\
$\Phi^{*0}_{s1}/\widehat{\Phi}^{*0}_{s1}$ & $\frac{1}{\sqrt{2}}(D^{*+}_{s0}D^-_{s1}+cD^+_{s1}D^{*-}_{s0})$ \\
\hline\hline
\end{tabular}
\end{table}

\subsection{Effective lagrangian}

With the help of the heavy quark symmetry and chiral symmetry, the
strong interaction between the P-wave $(0^+, 1^+)$ heavy doublet
reads
\begin{eqnarray}
\mathcal{L}&=& ig'Tr[S_b\gamma_\mu\gamma_5A^\mu_{ba}\bar{S}_a]+i\beta'Tr[S_bv^\mu(V_\mu-\rho_\mu)_{ba}\bar{S}_a]
\nonumber\\&&
+i\lambda'Tr[S_b\sigma^{\mu\nu}F_{\mu\nu}(\rho)_{ba}\bar{S}_a]+g'_{\sigma}Tr[S_a\sigma\bar{S}_a],
\end{eqnarray}
where $S$ represents the $(0^+,1^+)$ doublet. Its matrix
representation is
\begin{eqnarray}
S & = &
\frac{1}{2}(1+v\!\!\!\slash)[D^\mu_1\gamma_\mu\gamma_5-D^*_0], \\
\bar{S} & = & \gamma^0S^\dagger\gamma^0.
\end{eqnarray}
At the leading order, the axial vector field reads
\begin{eqnarray}
A^\mu_{ab}=\frac{1}{2}(\xi^\dagger\partial^\mu\xi-\xi\partial^\mu\xi^\dagger)_{ab}
=\frac{i}{f_\pi}\partial^\mu\mathcal{P}_{ab}+\ldots,
\end{eqnarray}
where
\begin{eqnarray}
\mathcal{P}=\left(\begin{array}{ccc}
\frac{\pi^0}{\sqrt{2}}+\frac{\eta}{\sqrt{6}} & \pi^+ & K^+ \\
\pi^- & -\frac{\pi^0}{\sqrt{2}}+\frac{\eta}{\sqrt{6}} & K^0 \\
K^- & \bar{K}^0 & -\frac{2\eta}{\sqrt{6}}
\end{array}\right).
\end{eqnarray}
$\rho^\mu_{ab}$ and $F^{\mu\nu}(\rho)_{ab}$ represent the vector
meson field and its strength tensor
\begin{eqnarray}\nonumber
\rho^\mu_{ab} & = & \frac{ig_v}{\sqrt{2}}\mathcal{V}^\mu_{ab},
\\ \nonumber
F^{\mu\nu}(\rho)_{ab} & = &
\partial^\mu\rho^\nu_{ab}-\partial^\nu\rho^\mu_{ab}+[\rho^\mu_{ab},\rho^\nu_{ab}]
\\ \nonumber & = &
\frac{ig_v}{\sqrt{2}}(\partial^\mu\mathcal{V}^\nu-\partial^\nu\mathcal{V}^\mu)_{ab}+\ldots,
\end{eqnarray}
where $g_v=m_\rho/f_\pi$ with $m_\rho=0.77$ GeV and $f_\pi=0.132$ GeV. $\mathcal{V}$ is the nonet vector meson matrices
\begin{eqnarray}
\mathcal{V}=\left(\begin{array}{ccc}
\frac{\rho^0}{\sqrt{2}}+\frac{\omega}{\sqrt{2}} & \rho^+ & K^{*+} \\
\rho^- & -\frac{\rho^0}{\sqrt{2}}+\frac{\omega}{\sqrt{2}}& K^{*0} \\
K^{*-} & \bar{K}^{*0} & \phi.
\end{array}\right).
\end{eqnarray}
Similarly the scalar field $\sigma$ represents the scalar nonet.
All the coupling constants $g',\beta',\lambda'$ and $g'_\sigma$
are real.

In our calculation we only need the effective lagrangian at the
tree level
\begin{eqnarray*}
\mathcal{L}_{D^*_0D^*_0\mathcal{V}} & = &
\sqrt{2}g_v\beta'v^\mu(\mathcal{V}_\mu)_{ba}D^*_{0b}D^{*\dagger}_{0a},
\\
\mathcal{L}_{D_1D_1\mathcal{V}} & = &
-\sqrt{2}g_v\beta'v^\mu(\mathcal{V}_\mu)_{ba}(D_{1b}\cdot
D^\dagger_{1a})
+2\sqrt{2}ig_v\lambda'(\partial_\mu\mathcal{V}_\nu
-\partial_\nu\mathcal{V}_\mu)_{ba}D^\mu_{1b}D^{\nu\dagger}_{1a},
\\
\mathcal{L}_{D^*_0D_1\mathcal{V}} & = &
-\sqrt{2}g_v\lambda'(\partial_\mu\mathcal{V}_\nu
-\partial_\nu\mathcal{V}_\mu)_{ba}\epsilon^{\alpha\mu\nu\beta}v_\beta
(D_{1b\alpha}D^{*\dagger}_{0a}+D^*_{0b}D^\dagger_{1a\alpha}),
\\
\mathcal{L}_{D_1D_1\mathcal{P}} & = &
\frac{2ig'}{f_\pi}\partial_\mu\mathcal{P}_{ba}D_{1b\alpha}D^\dagger_{1a\beta}\epsilon^{\alpha\mu\beta\nu}v_\nu,
\\
\mathcal{L}_{D^*_0D_1\mathcal{P}} & = &
-\frac{2g'}{f_\pi}\partial_\mu\mathcal{P}_{ba}(D^\mu_{1b}D^{*\dagger}_{0a}+D^*_{0b}D^{\mu\dagger}_{1a}),
\\
\mathcal{L}_{D^*_0D^*_0\sigma} & = & 2g'_\sigma D^*_{0a}
D^{*\dagger}_{0a}\sigma,
\\
\mathcal{L}_{D_1D_1\sigma} & = & -2g'_\sigma
(D_{1a}\cdot D^\dagger_{1a})\sigma.
\end{eqnarray*}
None of the coupling constants $g',\, \lambda',\, g'_\sigma$ are
known precisely although there exists some crude theoretical
estimation \cite{hpz}. We allow the
parameters involved in this work to vary around the values extracted
from the QCD sum rule approach (QSR).

\subsection{Derivation of the effective potential}

We follow Refs. \cite{liu1,liu2} to derive the effective potential
of the heavy molecular system. Interested readers may consult
Refs. \cite{liu1,liu2} for details. As usual, the monopole type
form factor (FF) is introduced at every interaction vertex in order
to account for the non-point-like structure effect of each
interacting particle and cure the singularity of the effective
potential.
\begin{eqnarray}
F(q)=\frac{\Lambda^2-m^2}{\Lambda^2-q^2}.
\end{eqnarray}
$\Lambda$ is the phenomenological cutoff parameter. Generally
$\Lambda$ is expected to be larger than the exchanged meson mass
and lies around 1-3 GeV.

The effective potential in the coordinate space reads
\begin{eqnarray}
\mathcal{V}(r)=\frac{1}{(2\pi)^3}\int
d\textbf{q}\mathcal{V}(\textbf{q})F(q)^2e^{-i\textbf{q}\cdot\textbf{r}},
\end{eqnarray}
\begin{eqnarray}
\frac{1}{\textbf{q}^2+m^2} & \longrightarrow & Y(\Lambda,m,r),
\\
\frac{\textbf{q}^2}{\textbf{q}^2+m^2} & \longrightarrow &
Z(\Lambda,m,r),
\end{eqnarray}
\begin{eqnarray}
Y(\Lambda,m,r) & = & \frac{1}{4\pi r}(e^{-mr}-e^{-\Lambda
r})-\frac{\xi^2}{8\pi\Lambda}e^{-\Lambda r},
\\
Z(\Lambda,m,r) & = & -\frac{1}{r^2}\frac{\partial}{\partial
r}(r^2\frac{\partial}{\partial r})Y(\Lambda,m,r)\nonumber\\
&=&-\frac{e^{-m r}m^2}{4\pi r}-\frac{e^{-\Lambda r}\xi^2}{4\pi r}+\frac{e^{-\Lambda r}\xi^2\Lambda}{8\pi}
+\frac{e^{-\Lambda r}\Lambda^2}{4\pi r},
\end{eqnarray}
with $\xi=\sqrt{\Lambda^2-m^2}$.

We collect the meson masses in Table \ref{sec2-tab3}.
\begin{table}[htb]
\centering \caption{The meson masses \cite{pdg}.}\label{sec2-tab3}
\begin{tabular}{cc|cc|cc}
\hline\hline
meson & mass (GeV) & mason & mess (GeV) & meson & mass (GeV) \\
\hline
$D^{*0}_0$ & $2.4$ & $D^{*\pm}_0$ & $2.4$ & $D^{*\pm}_{s0}$ & $2.317$ \\
$D^0_1$ & $2.42$ & $D^{\pm}_1$ & $2.42$ & $D^\pm_{s1}$ & $2.46$ \\
$\rho^0$ & $0.77$ & $\rho^\pm$ & $0.77$ & $\omega$ & $0.782$ \\
$\phi$ & $1.020$ & $\pi^0$ & 0.135 & $\pi^\pm$ & 0.140 \\
$\eta$ & $0.548$ & $\sigma$ & $0.66$ & $f_0(980)$ & 0.98\\
\hline\hline
\end{tabular}
\end{table}

\section{The $D^*_0-\bar{D}^*_0$ case}\label{sec3}

In the $D^*_0-\bar{D}^*_0$ case, the pseudoscalar meson exchange
is forbidden by parity and angular momentum conservation.
$\Phi^\pm_s$ and $\Phi^0_s(\bar{\Phi}^0_s)$ states don't exist.
For $\Phi^\pm,\Phi^0,\Phi^0_8,\Phi^0_{s1}$, the effective
potential reads
\begin{eqnarray}
V(r)^{\Phi^\pm,0}_{Total} & = &
-\frac{1}{4}g^2_v\beta'^2[-Y(\Lambda,m_\rho,r)+Y(\Lambda,m_\omega,r)]
-g'^2_\sigma Y(\Lambda,m_\sigma,r),
\end{eqnarray}
\begin{eqnarray}
V(r)^{\Phi^0_8}_{Total} & = &
-\frac{1}{4}g^2_v\beta'^2[3Y(\Lambda,m_\rho,r)+Y(\Lambda,m_\omega,r)]
-g'^2_\sigma Y(\Lambda,m_\sigma,r),
\end{eqnarray}
\begin{eqnarray}
V(r)^{\Phi^0_{s1}}_{Total}=-\frac{1}{2}g^2_v\beta'^2Y(\Lambda,m_\phi,r)-g'^2_\sigma Y(\Lambda,m_{f_0},r),
\end{eqnarray}

We use the MATSLISE package to solve Schr{\"o}dinger equation with
the effective potentials. We collect the variation of the binding
energy $E$ (in unit of MeV) and the root-mean-square radius $r$ (in
unit of fm) with the cutoff and the coupling constants in Table
\ref{sec3-tab1}.

As the coupling constants increase, the attraction becomes
stronger. The cutoff parameter reflects the non-point-like
structure of the interacting hadrons at each vertex. Its value is
the hadronic size. In this work we assume the "reasonable" cutoff
should be larger than the exchanged light meson mass and be around
1-3 GeV.

Simply for comparison, we also collect the numerical results for
the other possible molecular states in the same multiplet although
their experimental observation may be difficult because of the
broad width of the non-strange $(0^+, 1^+)$ charmed mesons.

\begin{table}[htb]
\centering \caption{The variation of the binding energy $E$ (in unit
of MeV) and the root-mean-square radius $r_{rms}$ (in unit of fm) with
the cutoff and the coupling constants for the $D^*_0-\bar{D}^*_0$
system.}\label{sec3-tab1}
$\beta'=0.84,\,g'_\sigma=0.761$ \\
\begin{tabular}{c|c|c|c}
\hline\hline
states & $\Lambda$ & $E$ & $r_{rms}$ \\
\hline
$\Phi$ & - & - & - \\
\hline
$\Phi_8$ & 1.6 & -8.1 & 1.28 \\
& 1.7 & -14.4 & 1.01 \\
& 1.8 & -21.9 & 0.86 \\
& 1.9 & -30.3 & 0.76 \\
\hline
$\Phi_{s1}$ & - & - & - \\
\hline\hline
\end{tabular}
\\
$\beta'=0.98,\,g'_\sigma=0.761$ \\
\begin{tabular}{c|c|c|c}
\hline\hline
states & $\Lambda$ & $E$ \\
\hline
$\Phi$ & - & - & - \\
\hline
$\Phi_8$ & 1.4 & -9.2 & 1.23 \\
& 1.45 & -14.0 & 1.04 \\
& 1.5 & -19.6 & 0.92 \\
& 1.55 & -25.8 & 0.83 \\
\hline
$\Phi_{s1}$ & - & - & - \\
\hline\hline
\end{tabular}
\\
$\beta'=1.12,\,g'_\sigma=0.761$ \\
\begin{tabular}{c|c|c|c}
\hline\hline
states & $\Lambda$ & $E$ & $r_{rms}$ \\
\hline
$\Phi$ & - & - & - \\
\hline
$\Phi_8$ & 1.25 & -6.9 & 1.39 \\
& 1.3 & -12.9 & 1.09 \\
& 1.35 & -20.4 & 0.92 \\
& 1.4 & -29.2 & 0.81 \\
\hline
$\Phi_{s1}$
& 3.0 & -17.8 & 0.87 \\
& 3.2 & -25.6 & 0.75 \\
& 3.4 & -34.1 & 0.66 \\
\hline\hline
\end{tabular}

\end{table}

\section{The $D_1-\bar{D}_1$ case}\label{sec4}

The effective potential of the $\Phi^{**\pm}$, $\Phi^{**0}$,
$\Phi^{**0}_8$, $\Phi^{**\pm}_s$, $\Phi^{**0}_s$,
$\Phi^{**0}_{s1}$ systems reads
\begin{eqnarray}
V(r)^{\Phi^{**\pm,0}[J]}_{Total} & =&
-\frac{1}{4}g^2_v\beta'^2\mathcal{C}(J)[-Y(\Lambda,m_\rho,r)+Y(\Lambda,m_\omega,r)]
-\lambda'^2g^2_v\mathcal{B}(J)[-Z(\Lambda,m_\rho,r)+Z(\Lambda,m_\omega,r)]
\nonumber\\ &&
+\frac{g'^2}{2f_\pi^2}\mathcal{A}(J)[-Z(\Lambda,m_\pi,r))+\frac{1}{3}Z(\Lambda,m_\eta,r))]
-g'^2_\sigma\mathcal{C}(J)Y(\Lambda,m_\sigma,r),
\end{eqnarray}
\begin{eqnarray}
V(r)^{\Phi^{**0}_8[J]}_{Total}  & =&
-\frac{1}{4}g^2_v\beta'^2\mathcal{C}(J)[3Y(\Lambda,m_\rho,r)+Y(\Lambda,m_\omega,r)]
-\lambda'^2g^2_v\mathcal{B}(J)[3Z(\Lambda,m_\rho,r)+Z(\Lambda,m_\omega,r)]
\nonumber\\ &&
+\frac{g'^2}{2f_\pi^2}\mathcal{A}(J)[3Z(\Lambda,m_\pi,r))+\frac{1}{3}Z(\Lambda,m_\eta,r))]
-g'^2_\sigma\mathcal{C}(J)Y(\Lambda,m_\sigma,r),
\end{eqnarray}
\begin{eqnarray}
V(r)^{\Phi^{**\pm}_s[J]}_{Total} & = &
-\frac{g'^2}{3f_\pi^2}\mathcal{A}(J)Z(\Lambda,m_\eta,r),
\end{eqnarray}
\begin{eqnarray}
V(r)^{\Phi^{**0}_s/\bar{\Phi}{}^{**0}_s[J]}_{Total} & = &
-\frac{g'^2}{3f_\pi^2}\mathcal{A}(J)Z(\Lambda,m_\eta,r),
\end{eqnarray}
\begin{eqnarray}
V(r)^{\Phi^{**0}_{s1}[J]}_{Total} & = &
-\frac{1}{2}g^2_v\beta'^2\mathcal{C}(J)Y(\Lambda,m_\phi,r)
-2\lambda'^2g^2_v\mathcal{B}(J)Z(\Lambda,m_\phi,r)
\nonumber\\ &  &
+\frac{2g'^2}{3f_\pi^2}\mathcal{A}(J)Z(\Lambda,m_\eta,r)
-g'^2_\sigma\mathcal{C}(J)Y(\Lambda,m_{f_0},r),
\end{eqnarray}
where $\mathcal{A}(J)$, $\mathcal{B}(J)$ and $\mathcal{C}(J)$
denote
\begin{eqnarray}
\mathcal{A}(J) & \equiv &
\sum_{\lambda_1\lambda_2\lambda_3\lambda_4}\langle1\lambda_1;1\lambda_2|J,m\rangle\langle1\lambda_3;1\lambda_4|J,m\rangle
\frac{1}{\vec{q}^2}
[\vec{\epsilon}^{\lambda1}_1\cdot(\vec{q}\times\vec{\epsilon}^{\lambda3*}_3)\vec{\epsilon}^{\lambda2}_2\cdot(\vec{q}\times\vec{\epsilon}^{\lambda4*}_4)],
\\
\mathcal{B}(J) & \equiv &
\sum_{\lambda_1\lambda_2\lambda_3\lambda_4}\langle1\lambda_1;1\lambda_2|J,m\rangle\langle1\lambda_3;1\lambda_4|J,m\rangle
\frac{1}{\vec{q}^2}
[(\vec{\epsilon}^{\lambda1}_1\cdot\vec{q})(\vec{\epsilon}^{\lambda2}_2\cdot\vec{q})(\vec{\epsilon}^{\lambda3*}_3\cdot\vec{\epsilon}^{\lambda4*}_4)+(c.t.s)],
\\
\mathcal{C}(J) & \equiv &
\sum_{\lambda_1\lambda_2\lambda_3\lambda_4}\langle1\lambda_1;1\lambda_2|J,m\rangle\langle1\lambda_3;1\lambda_4|J,m\rangle
[(\vec{\epsilon}^{\lambda1}_1\cdot\vec{\epsilon}^{\lambda3*}_3)(\vec{\epsilon}^{\lambda2}_2\cdot\vec{\epsilon}^{\lambda4*}_4)],
\end{eqnarray}
$\vec{\epsilon}_1,\vec{\epsilon}_2,\vec{\epsilon}_3,\vec{\epsilon}_4$
are the polarizations of the initial and final states. $c.t.s$
denotes
\begin{eqnarray}
c.t.s&=&(\vec{\epsilon}^{\lambda3*}_3\cdot\vec{q})(\vec{\epsilon}^{\lambda4*}_4\cdot\vec{q})(\vec{\epsilon}^{\lambda1}_1\cdot\vec{\epsilon}^{\lambda2}_2)
-(\vec{\epsilon}^{\lambda1}_1\cdot\vec{q})(\vec{\epsilon}^{\lambda4*}_4\cdot\vec{q})(\vec{\epsilon}^{\lambda2}_2\cdot\vec{\epsilon}^{\lambda3*}_3)
\nonumber \\
&&-(\vec{\epsilon}^{\lambda2}_2\cdot\vec{q})(\vec{\epsilon}^{\lambda3*}_3\cdot\vec{q})(\vec{\epsilon}^{\lambda1}_1\cdot\vec{\epsilon}^{\lambda4*}_4),
\nonumber
\end{eqnarray}
The values of $\mathcal{A}(J)$, $\mathcal{B}(J)$ and
$\mathcal{C}(J)$ with different quantum numbers are listed in
Table \ref{sec4-tab1}.
\begin{table}[htb]
\centering \caption{The values of $\mathcal{A}(J)$,
$\mathcal{B}(J)$ and $\mathcal{C}(J)$ with different quantum
numbers.}\label{sec4-tab1}
\begin{tabular}{c|cccc}
\hline\hline
$J$ & $\mathcal{A}(J)$ & $\mathcal{B}(J)$ & $\mathcal{C}(J)$ \\
\hline
$0$ & $\frac{2}{3}$ & $\frac{4}{3}$ & $1$ \\
$1$ & $\frac{1}{3}$ & $\frac{2}{3}$ & $1$ \\
$2$ & $-\frac{1}{3}$ & $-\frac{2}{3}$ & $1$ \\
\hline\hline
\end{tabular}
\end{table}

We collect the variation of the binding energy $E$ and the
root-mean-square radius $r$ with the cutoff and the coupling
constants in Tables \ref{sec4-tab2}-\ref{sec4-tab3}.
With the pseudoscalar meson
exchange force alone and $g^\prime=0.80$, there exists an
isovector $\Phi^{**}$ with $J^P=0^+$, a $\Phi_8^{**}$ state with
$J^P=2^+$, an isoscalar $\Phi_s^{**}$ state with $J^P=0^+$ and an
isoscalar $\Phi_{s1}^{**}$ state with $J^P=2^+$. Increasing
$g^\prime$ to 1.06, we can find the bound state solution for
$\Phi^{**}$ with $J^P=1^+$ and $\Phi_s^{**}$ with $J^P=1^+$
besides the above mentioned bound states. With $g^\prime =1.32$,
the above bound states still exist. We notice that corresponding
cutoff $\Lambda$ becomes smaller with the larger $g^\prime$.

Including all the exchange meson
contributions, we list the numerical results in Table
\ref{sec4-tab3}. From Tables \ref{sec4-tab2} and \ref{sec4-tab3}
we note that the pseudoscalar meson exchange potential is dominant
in the total effective potential. Thus, it is reasonable to
consider pseudoscalar meson exchange potential only when studying
whether there exists a bound state solution for the $D_1-\bar D_1$
case. Meanwhile, for the $B_1-\bar{B}_1$ system, we list the
results in Table \ref{b1b1}.

\section{The $D^*_0-\bar{D}_1$ case}\label{sec5}

In the $D^*_0-\bar{D}_1$ case, there are both direct and crossed
scattering channels in the derivation of the effective potential
in the momentum space. In the crossed channel, the mass difference
$q_0$ between the initial and final states (i.e., $D^*_0$ and
$D_1$) should be kept. We introduce
\begin{eqnarray}
& \mu_m=\sqrt{m^2-q^2_0}, \nonumber \\
& \alpha=\sqrt{\Lambda^2-q^2_0}, \nonumber
\end{eqnarray}
where the subscript "$m$" denotes the exchanged meson. Accordingly,
\begin{eqnarray}
F(q)=\frac{\Lambda^2-m^2}{\Lambda^2-q^2}=\frac{\alpha^2-\mu^2}{\alpha^2+\textbf{q}^2}\;
. \nonumber
\end{eqnarray}
After the Fourier transformation
\begin{eqnarray}
\frac{1}{q^2-m^2}=\frac{1}{q^2_0-\textbf{q}^2-m^2} &
\longrightarrow & -Y(\alpha,\mu,r),\\
\frac{\textbf{q}^2}{q^2-m^2}=\frac{\textbf{q}^2}{q^2_0-\textbf{q}^2-m^2}
& \longrightarrow & -Z(\alpha,\mu,r),
\end{eqnarray}
the effective potential for the $\Phi^{*\pm}$,
$\widehat{\Phi}^{*\pm}$, $\Phi^{*0}$, $\widehat{\Phi}^{*0}$,
$\Phi^{*0}_8$, $\widehat{\Phi}^{*0}_8$, $\Phi^{*\pm}_s$,
$\widehat{\Phi}^{*\pm}_s$, $ \Phi^{*0}_s$,
$\widehat{\Phi}^{*0}_s$, $\Phi^{*0}_{s1}$,
$\widehat{\Phi}^{*0}_{s1}$ systems read as
\begin{eqnarray}
V(r)^{\Phi^{*\pm,0}/\widehat{\Phi}^{*\pm,0}}_{Total}& =&
-\frac{1}{4}g^2_v\beta'^2[-Y(\Lambda,m_\rho,r)+Y(\Lambda,m_\omega,r)]
+c\Big\{\frac{2}{3}g^2_v\lambda'^2[-Z(\alpha,\mu_\rho,r)+Z(\alpha,\mu_\omega,r)]
\nonumber\\ &  &\quad
-\frac{1}{6}\frac{g'^2}{f^2_\pi}[-Z(\alpha,\mu_\pi,r)+\frac{1}{3}Z(\alpha,\mu_\eta,r)]\Big\}
 -g'^2_\sigma Y(\Lambda,m_\sigma,r),
\end{eqnarray}
\begin{eqnarray}
V(r)^{\Phi^{*0}_8/\widehat{\Phi}^{*0}_8}_{Total}  & = &
-\frac{1}{4}g^2_v\beta'^2[3Y(\Lambda,m_\rho,r)+Y(\Lambda,m_\omega,r)]
+c\Big\{\frac{2}{3}g^2_v\lambda'^2[3Z(\alpha,\mu_\rho,r)+Z(\alpha,\mu_\omega,r)]
\nonumber\\ &  &
-\frac{1}{6}\frac{g'^2}{f^2_\pi}[3Z(\alpha,\mu_\pi,r)+\frac{1}{3}Z(\alpha,\mu_\eta,r)]\Big\}
 -g'^2_\sigma Y(\Lambda,m_\sigma,r),
\end{eqnarray}
\begin{eqnarray}
V(r)^{\Phi^{*\pm}_s/\widehat{\Phi}^{*\pm}_s}_{Total}  & = &
c\frac{1}{9}\frac{g'^2}{f^2_\pi}Z(\alpha,\mu_\eta,r),
\end{eqnarray}
\begin{eqnarray}
V(r)^{\Phi^{*0}_s,\bar{\Phi}^{*0}_s/\widehat{\Phi}^{*0}_s,\widehat{\bar{\Phi}}{}^{*0}_s}_{Total} & = &
c\frac{1}{9}\frac{g'^2}{f^2_\pi}Z(\alpha,\mu_\eta,r),
\end{eqnarray}
\begin{eqnarray}
 V(r)^{\Phi^{*0}_{s1}/\widehat{\Phi}^{*0}_{s1}}_{Total} & &=
-\frac{1}{2}g^2_v\beta'^2Y(\Lambda,m_\phi,r)
+c\Big\{\frac{4}{3}g^2_v\lambda'^2Z(\alpha,\mu_\phi,r)
-\frac{2}{9}\frac{g'^2}{f^2_\pi}Z(\alpha,\mu_\eta,r)\Big\}
 -g'^2_\sigma Y(\Lambda,m_{f_0},r).
\end{eqnarray}

The $D^*_0-\bar{D}_1$ system is very similar to the $D-\bar
D^\ast$ case and is particularly interesting since $X(3872)$ is
often speculated to be a $D-\bar D^\ast$ molecular candidate. The
only difference is that both components in the $D^*_0-\bar{D}_1$
system are extremely narrow P-wave states. We first focus on the
pseudoscalar meson exchange, which is repulsive for the
$\Phi^{*0}_{s1}$ state with negative charge parity. The
$J^{PC}=1^{++}$ $\Phi^{*0}_{s1}$ state appears
as shown in Tables
\ref{sec5-tab1}-\ref{sec5-tab2}. By comparing the result listed in
Table \ref{sec5-tab1} and that in Table \ref{sec5-tab2}, one
notices that the pseudoscalar meson exchange is dominant in the
$D^*_0-\bar{D}_1$ system, which shows that it is reasonable to
consider the pseudoscalar meson exchange potential only when we
investigate whether there exists the bound state solution for the
$D^*_0-\bar{D}_1$ system. The result for the $B^*_0-\bar{B}_1$
system corresponds to the pseudoscalar meson exchange only.

\section{Conclusion}\label{sec6}

Both $D_{s0}(2317)$ and $D_{s1}(2460)$ lie below the $DK$ and
$D^\ast K$ threshold respectively. They are extremely narrow. The
possible molecular states composed of the $D_{s0}(2317)$ and
$D_{s1}(2460)$ may be observable experimentally if they really
exist. In this work we have studied such systems carefully. As a
byproduct, we collect the numerical results for the
bottomonium-like analogue in the appendix.

One should be cautious that our numerical results are quite
sensitive to the values of the hadronic coupling constants. The
values are larger than (or around the upper bound of) those derived
from the crude estimate with the light cone QCD sum rule approach
\cite{hpz}. Future lattice QCD simulations may help extract these
coupling constants more precisely. Since the hadronic coupling
constants are not known well, we allow them to vary.
As shown in the numerical result,
the binding energy is also sensitive to the value of the cutoff
introduced in the form factor. Thus, further study and improvement
of the potential model are still desirable.

Here, we need to emphasize that a
monopole form factor is introduced in the numerical calculation of
this work. In fact, there are many types of form factor, such as
the dipole form factor. When taking the other type of the form
factor, the qualitative conclusion keeps the same as that obtained
in this work. Both the form factor and the cutoff are necessary and
important for the hadronic system since the components are not
point-like particles. They are hadrons with internal structure. When
dealing with the loosely bound heavy molecular states, only the
relatively soft degree of freedom is expected to play the dominant
role. The exchanged soft mesons should not "see" the quark/gluon
structure of the heavy meson. That's the physical meaning of the
form factor and the cutoff.

So long as these couplings are big enough, there may even appear
deeply bound states including radial and orbital excitations.
However they are no more the "conventional" molecular states,
which are loosely bound with a typical binding energy around
several to several tens MeV and a radius around 1.5-3 fm.
Therefore we do not list numerical results for the deeply bound
cases in this work.

From our calculation there may exist two loosely bound $0^{++}$
charmonium-like states, the first of which is composed of the
$D_{s0}{\bar D}_{s0}$ pair and lies around 4.61 GeV. The other one
is around 4.9 GeV and composed of the $D_{s1}$ and ${\bar D}_{s1 }$
pair. There exists the $2^{++}$ $D_{s1}\bar{D}_{s1}$ state, which
lies around 4.9 GeV. The $1^{++}$ state around 4.75 GeV is composed
of the $D_{s0}$ and ${\bar D}_{s1 }$ pair. This state is very
interesting because of its similarity to $X(3872)$.

The dominant decay modes of the above states are the open-charm
modes $D_s^{(\ast)} {\bar D}_s^{(\ast)}$. The other characteristic
decay modes are the hidden-charm modes $J/\psi \phi$, $\eta_c
\eta'$, $\eta_c f_0(980)$, $\chi_{cJ}\eta'$, $\chi_{cJ} f_0(980)$,
$\psi' \phi$, $\psi'' \phi$, $\eta_c(2S) \eta'$, etc. for the
possible $C=+$ molecular states. One may easily exhaust the possible
final states according to the $C/P$ parity and angular momentum
conservation and kinematical considerations. These states may be
significantly narrower than the conventional charmonium around the
same mass region because of their molecular nature. However, their
widths should be larger than those of $X(3872)$ due to much larger
phase space and more decay modes.

These states might be produced from $B$ or $B_s$ decays if
kinematically allowed. Those states with $J^{PC}=0^{++}, 2^{++}$
may be produced from the two photon fusion process at the $e^+e^-$
collider at $B$ factories. The other possible facilities to look for
them are RHIC, Tevatron and LHCb. Investigations of these states
may help us understand the puzzling $X(3872)$ state.

\begin{table}[htb]
\centering \caption{The variation of the binding energy $E$ (in unit
of MeV) and the root-mean-square radius $r_{rms}$ (in unit of fm) with
the cutoff and the coupling constant for the $D_1-\bar{D}_1$
system when only the pseudoscalar meson exchange is
considered. Here, we scan the cutoff range $\Lambda\leq 3.2$ GeV. }\label{sec4-tab2}
$g'=0.80,\,\beta'=0,\,\lambda'=0,\,g'_\sigma=0$ \\
\begin{tabular}{c|ccc|ccc|ccccc}
\hline\hline
      & &$J^P=0^+$ && &$J^P=1^+$& && $J^P=2^+$&\\ \cline{2-10}
state & $\Lambda$ & $E$ & $r_{rms}$& $\Lambda$ & $E$ & $r_{rms}$& $\Lambda$ & $E$ & $r_{rms}$\\\hline
$\Phi^{**}$& 1.9 & -5.4 & 1.33 &- &- &- &- &- &-\\
           & 2.0 & -10.6& 0.98 &- &- &- &- &- &-\\
           & 2.1 & -17.9& 0.78 &- &- &- &- &- &-\\
           & 2.2 & -27.9& 0.65 &- &- &- &- &- &-\\\hline
$\Phi_8^{**}$&-&-&-&-&-&-&1.1&-4.5 &1.49\\
             &-&-&-&-&-&-&1.2&-13.6&0.93\\
             &-&-&-&-&-&-&1.3&-28.9&0.69\\\hline
$\Phi_s^{**}$& 2.7 & -6.58 & 1.13 &- &- &- &- &- &-\\
             & 2.8 & -18.5 &  0.70 &- &- &- &- &- &-\\
             & 2.9 & -36.4 &  0.52 &- &- &- &- &- &-\\\hline
$\Phi_{s1}^{**}$&-&-&-&-&-&-&2.7&-7.8 &1.03\\
                &-&-&-&-&-&-&2.8&-20.5&0.66\\
                &-&-&-&-&-&-&2.9&-39.2&0.50\\\hline\hline

\end{tabular}
\\
$g'=1.06,\,\beta'=0,\,\lambda'=0,\,g'_\sigma=0$ \\

\begin{tabular}{c|ccc|ccc|ccccc}
\hline\hline
      & &$J^P=0^+$ && &$J^P=1^+$& && $J^P=2^+$&\\ \cline{2-10}
state & $\Lambda$ & $E$ & $r_{rms}$& $\Lambda$ & $E$ & $r_{rms}$& $\Lambda$ & $E$ & $r_{rms}$\\\hline
$\Phi^{**}$& 1.1 & -4.0 & 1.58 & 2.1 &-3.4 &1.62 &- &- &-\\
           & 1.2 & -9.9&  1.08 & 2.2 &-7.4 &1.14 &- &- &-\\
           & 1.3 & -19.0& 0.82 & 2.3 &-13.3 &0.88 &- &- &-\\
           & 1.4 & -31.9& 0.67 & 2.4 &-21.3 &0.72 &- &- &-\\\hline
$\Phi_8^{**}$&-&-&-&-&-&-&0.75&-3.6&1.72\\
             &-&-&-&-&-&-&0.8&-8.0&1.23\\
             &-&-&-&-&-&-&0.9&-23.6&0.81\\\hline
$\Phi_s^{**}$& 1.9 & -4.3 & 1.39 &3.0 &-11.7 &0.86 &- &- &-\\
             & 2.0 & -18.9 & 0.71 &3.1 &-25.6 &0.60 &- &- &-\\
             & 2.1 & -43.4 &  0.50 &3.2 &-44.8 &0.47 &- &- &-\\\hline
$\Phi_{s1}^{**}$&-&-&-&-&-&-&1.9&-5.1 &1.28\\
                &-&-&-&-&-&-&2.0&-20.5&0.68\\
                &-&-&-&-&-&-&2.1&-45.7&0.49\\\hline\hline

\end{tabular}
\\
$g'=1.32,\,\beta'=0,\,\lambda'=0,\,g'_\sigma=0$ \\

\begin{tabular}{c|ccc|ccc|ccccc}
\hline\hline
      & &$J^P=0^+$ && &$J^P=1^+$& && $J^P=2^+$&\\ \cline{2-10}
state & $\Lambda$ & $E$ & $r_{rms}$& $\Lambda$ & $E$ & $r_{rms}$& $\Lambda$ & $E$ & $r_{rms}$\\\hline
$\Phi^{**}$& 0.75 & -3.0 & 1.86 & 1.35 &-2.7 &1.85 &- &- &-\\
           & 0.85 & -10.3&  1.11 & 1.50 &-10.0 &1.04 &- &- &-\\
           & 0.95 & -22.8& 0.82 & 1.65 &-23.1 &0.73 &- &- &-\\
           & 1.05 & -41.3&  0.65 & 1.75 &-35.9 &0.61 &- &- &-\\\hline
$\Phi_8^{**}$&-&-&-&-&-&-&0.65&-12.0&1.11\\
             &-&-&-&-&-&-&0.70&-22.0&0.88\\
             &-&-&-&-&-&-&0.75&-35.9&0.73\\\hline
$\Phi_s^{**}$& 1.55 & -8.4 & 1.03 &2.25 &-9.7&0.95 &- &- &-\\
             & 1.60 & -18.5 & 0.73 &2.30 &-16.9 &0.74 &- &- &-\\
             & 1.65& -32.4 &  0.58 &2.40 &-37.1 &0.53 &- &- &-\\\hline
$\Phi_{s1}^{**}$&-&-&-&-&-&-&1.50&-2.6 &1.78\\
                &-&-&-&-&-&-&1.55&-9.3&0.98\\
                &-&-&-&-&-&-&1.60&-19.8&0.71\\\hline\hline

\end{tabular}
\end{table}

\begin{table}[htb]
\centering \caption{The variation of the binding energy $E$ (in unit
of MeV) and the root-mean-square radius $r_{rms}$ (in unit of fm) with
the cutoff and the coupling constants for the $D_1-\bar{D}_1$
system. Here, we scan the cutoff range $\Lambda\leq 3.2$ GeV.}\label{sec4-tab3}

$g'=0.80,\,\beta'=0.84,\,\lambda'=0.42,\,g'_\sigma=0.761$ \\
\begin{tabular}{c|ccc|ccc|ccccc}
\hline\hline
      & &$J^P=0^+$ && &$J^P=1^+$& && $J^P=2^+$&\\ \cline{2-10}
state & $\Lambda$ & $E$ & $r_{rms}$& $\Lambda$ & $E$ & $r_{rms}$& $\Lambda$ & $E$ & $r_{rms}$\\\hline
$\Phi^{**}$& 1.9 & -7.1 & 1.18 &- &- &- &- &- &-\\
           & 2.0 & -13.0& 0.91 &- &- &- &- &- &-\\
           & 2.1 & -21.0& 0.74 &- &- &- &- &- &-\\
           & 2.2 & -31.7& 0.62 &- &- &- &- &- &-\\\hline
$\Phi_8^{**}$&-&-&-&-&-&-&1.0&-5.7 &1.41\\
             &-&-&-&-&-&-&1.05&-11.6&1.06\\
             &-&-&-&-&-&-&1.1&-19.5&0.87\\\hline
$\Phi_s^{**}$& 2.7 & -6.58 & 1.13 &- &- &- &- &- &-\\
             & 2.8 & -18.5 &  0.70 &- &- &- &- &- &-\\
             & 2.9 & -36.4 &  0.52 &- &- &- &- &- &-\\\hline
$\Phi_{s1}^{**}$&-&-&-&-&-&-&2.6&-3.2 &1.70\\
                &-&-&-&-&-&-&2.9&-10.0&1.03\\
                &-&-&-&-&-&-&3.2&-18.8&0.79\\
                &-&-&-&-&-&-&3.5&-28.6&0.66\\\hline\hline

\end{tabular}
\\
$g'=1.06,\,\beta'=0.98,\,\lambda'=0.49,\,g'_\sigma=0.761$ \\
\begin{tabular}{c|ccc|ccc|ccccc}
\hline\hline
      & &$J^P=0^+$ && &$J^P=1^+$& && $J^P=2^+$&\\ \cline{2-10}
state & $\Lambda$ & $E$ & $r_{rms}$& $\Lambda$ & $E$ & $r_{rms}$& $\Lambda$ & $E$ & $r_{rms}$\\\hline
$\Phi^{**}$& 1.1 & -4.1 & 1.57 & 2.1 &-5.1 &1.35 &- &- &-\\
           & 1.2 & -10.1&  1.07 & 2.2 &-9.9 &1.01 &- &- &-\\
           & 1.3 & -19.3& 0.82 & 2.3 &-16.6 &0.81 &- &- &-\\
           & 1.4 & -32.4& 0.66 & 2.4 &-25.5 &0.67 &- &- &-\\\hline
$\Phi_8^{**}$&-&-&-&-&-&-&0.8&-8.4&1.21\\
             &-&-&-&-&-&-&0.85&-16.9&0.93\\
             &-&-&-&-&-&-&0.9&-29.7&0.76\\\hline
$\Phi_s^{**}$& 1.9 & -4.3 & 1.39 &2.9 &-3.1&1.61 &- &- &-\\
             & 1.95 & -10.4 & 0.92 &3.0 &-11.7&0.86 &- &- &-\\
             & 2.0 & -18.9 & 0.71 &3.1 &-25.6 &0.60 &- &- &-\\
             & 2.05 & -29.9 &  0.58 &3.15 &-34.5 &0.53 &- &- &-\\\hline
$\Phi_{s1}^{**}$&-&-&-&-&-&-&1.75&-3.6 &1.60\\
                &-&-&-&-&-&-&1.8&-6.9&1.20\\
                &-&-&-&-&-&-&1.9&-16.1&0.83\\
                &-&-&-&-&-&-&2.0&-28.1&0.66\\\hline\hline

\end{tabular}
\\
$g'=1.32,\,\beta'=1.12,\,\lambda'=0.56,\,g'_\sigma=0.761$ \\
\begin{tabular}{c|ccc|ccc|ccccc}
\hline\hline
      & &$J^P=0^+$ && &$J^P=1^+$& && $J^P=2^+$&\\ \cline{2-10}
state & $\Lambda$ & $E$ & $r_{rms}$& $\Lambda$ & $E$ & $r_{rms}$& $\Lambda$ & $E$ & $r_{rms}$\\\hline
$\Phi^{**}$& 0.8 & -6.1 & 1.38 & 1.4 &-4.9 &1.42 &- &- &-\\
           & 0.85 & -10.3&  1.12 & 1.5 &-10.6 &1.02 &- &- &-\\
           & 0.9 & -15.7& 0.95 & 1.6 &-19.0 &0.80 &- &- &-\\
           & 0.95 & -22.5&  0.82 & 1.7 &-30.5 &0.66 &- &- &-\\\hline
$\Phi_8^{**}$&-&-&-&-&-&-&0.8&-55.1&0.62\\
             &-&-&-&-&-&-&0.9&-118.0&0.48\\
             &-&-&-&-&-&-&1.0&-213.9&0.39\\\hline
$\Phi_s^{**}$& 1.55 & -8.4 & 1.03 &2.2 &-4.4&1.42 &- &- &-\\
             & 1.60 & -18.7 & 0.73 &2.25 &-9.7 &0.95 &- &- &-\\
             & 1.65& -32.4 &  0.58 &2.3 &-16.9 &0.74 &- &- &-\\\hline
$\Phi_{s1}^{**}$&-&-&-&-&-&-&1.45&-5.4 &1.32\\
                &-&-&-&-&-&-&1.50&-12.5 &0.92\\
                &-&-&-&-&-&-&1.55&-22.0&0.73\\
                &-&-&-&-&-&-&1.60&-33.7&0.62\\\hline\hline

\end{tabular}
\end{table}

\begin{table}[htb]
\centering \caption{The variation of the binding energy $E$ (in unit
of MeV) and the root-mean-square radius $r_{rms}$ (in unit of fm) with
the cutoff and the coupling constant for the $D^*_0-\bar{D}_1$
system when only the pseudoscalar meson exchange is
considered.}\label{sec5-tab1}

$g'=0.80,\,\beta'=0,\,\lambda'=0,\,g'_\sigma=0$ \\
\begin{tabular}{c|ccc|ccc}
\hline\hline
&& $c=+1$ &&& $c=-1$ \\
\hline
states & $\Lambda$ & $E$ & $r_{rms}$ & $\Lambda$ & $E$ & $r_{rms}$ \\
\hline
$\Phi^{*}$ & - & - & - & - & - & - \\
\hline
$\Phi^{*}_8$ & 1.1 & -4.4 & 1.51 & - & - & - \\
             & 1.2 & -13.4 &0.94 & - & - & - \\
             & 1.3 & -28.6 & 0.69 & - & - & - \\
             & 1.4 & -51.3 & 0.55 & - & - & - \\
\hline
$\Phi^{*}_{s1}$ & 2.7 & -5.2  & 1.27 & - & - & - \\
                & 2.8 & -16.0 & 0.75 &&& \\
                & 2.85 &-23.6 & 0.63 &&& \\
                & 2.9 & -32.6 & 0.55 &&& \\
\hline\hline
\end{tabular}
\\
$g'=1.06,\beta'=0,\,\lambda'=0,\,g'_\sigma=0$ \\
\begin{tabular}{c|ccc|ccc}
\hline\hline
&& $c=+1$ &&& $c=-1$ \\
\hline
states & $\Lambda$ & $E$ & $r_{rms}$ & $\Lambda$ & $E$ & $r_{rms}$ \\
\hline
$\Phi^{*}$ & - & -& -& 2.2 & -7.1 & 1.17 \\
           & - & -& -& 2.3 & -12.8 &0.90  \\
           & - & -& -& 2.4 & -20.7 &0.73 \\
           & - & -& -& 2.5 & -30.9 &0.61 \\
\hline
$\Phi^{*}_8$ & 0.8 & -8.0 & 1.24 & - & - & - \\
             & 0.85 & -14.5 & 0.98 & - & - & - \\
             & 0.9 & -23.5 & 0.81 & - & - & - \\
             & 0.95 & -35.3 &0.69 & - & - & - \\
\hline
$\Phi^{*}_s$ & - & - & -& 3 & -9.0 & 0.98  \\
             & - & - & -& 3.05 & -14.5 & 0.78  \\
             & - & - & -& 3.1 & -21.4 & 0.66  \\
             & - & - & - & 3.15 & -29.6 &0.57 \\
\hline
$\Phi^{*}_{s1}$ & 1.9 & -4.2 & 1.43 & - & - & - \\
                & 1.95& -10.1 &0.95 &-&-&- \\
                & 2.0 & -18.3 &0.73 &-&-&- \\
                & 2.05 &-29.0 &0.60 &-&-&- \\
\hline\hline
\end{tabular}
\\
$g'=1.32,\,\beta'=0,\,\lambda'=0,\,g'_\sigma=0$ \\
\begin{tabular}{c|ccc|ccc}
\hline\hline
&& $c=+1$ &&& $c=-1$ \\
\hline
states & $\Lambda$ & $E$ & $r_{rms}$ & $\Lambda$ & $E$ & $r_{rms}$ \\
\hline
$\Phi^{*}$ & - & -& -& 1.5 & -9.7 & 1.06 \\
           & - & -& -& 1.6 & -17.7 &0.82  \\
           & - & -& -& 1.7 & -28.6 &0.67 \\
\hline
$\Phi^{*}_8$ & 0.8 & -54.2 & 0.63 & - & - & - \\
             & 0.825 & -65.4 & 0.58 & - & - & - \\
             & 0.85 & -77.9 & 0.55 & - & - & - \\
             & 0.875 & -92.0 &0.51 & - & - & - \\
\hline
$\Phi^{*}_s$ & - & - & -& 2.25 & -7.9 & 1.05  \\
             & - & - & - & 2.3 & -14.5 & 0.80 \\
             & - & - & -& 2.35 & -23.0 & 0.65  \\
             & - & - & -& 2.4 & -33.5 &0.55  \\
\hline
$\Phi^{*}_{s1}$ & 1.55 & -8.9 & 1.02 & - & - & - \\
                & 1.6& -19.1 &0.73 &-&-&- \\
                & 1.65 & -32.9 &0.59 &-&-&- \\
                & 1.7 &-50.5 &0.49 &-&-&- \\
\hline\hline
\end{tabular}

\end{table}

\begin{table}[htb]
\centering \caption{The variation of the binding energy $E$ (in unit
of MeV) and the root-mean-square radius $r_{rms}$ (in unit of fm) with
the cutoff and the coupling constants for the $D^*_0-\bar{D}_1$
system.}\label{sec5-tab2}

$g'=0.80,\,\beta'=0.84,\,\lambda'=0.42,\,g'_\sigma=0.761$ \\
\begin{tabular}{c|ccc|ccc}
\hline\hline
&& $c=+1$ &&& $c=-1$ \\
\hline
states & $\Lambda$ & $E$ & $r_{rms}$ & $\Lambda$ & $E$ & $r_{rms}$ \\
\hline
$\Phi^{*}$ & - & - & - & - & - & - \\
\hline
$\Phi^{*}_8$ & 1.0 & -5.6 & 1.42 & - & - & - \\
             & 1.05 & -11.5 &1.07 & - & - & - \\
             & 1.1 & -19.3 & 0.88 & - & - & - \\
             & 1.15 & -28.8 & 0.76 & - & - & - \\
\hline
$\Phi^{*}_{s1}$ & 2.7 & -4.0  & 1.56& - & - & - \\
                & 3.0 & -10.8 & 1.01 &&& \\
                & 3.3 &-19.1 & 0.79 &&& \\
                & 3.6 & -28.3 & 0.67 &&& \\
\hline\hline
\end{tabular}
\\
$g'=1.06,\beta'=0.98,\lambda'=0.49,g'_\sigma=0.761$ \\
\begin{tabular}{c|ccc|ccc}
\hline\hline
&& $c=+1$ &&& $c=-1$ \\
\hline
states & $\Lambda$ & $E$ & $r_{rms}$ & $\Lambda$ & $E$ & $r_{rms}$ \\
\hline
$\Phi^{*}$ & - & -& -& 2.1 & -4.9 &1.39 \\
           & - & -& -& 2.2 & -9.6 &1.03  \\
           & - & -& -& 2.3 & -16.1 &0.82 \\
           & - & -& -& 2.4 & -24.8 &0.68 \\
\hline
$\Phi^{*}_8$ & 0.8 & -8.3 & 1.22 & - & - & - \\
             & 0.85 & -16.7 & 0.93 & - & - & - \\
             & 0.9 & -29.5 & 0.76 & - & - & - \\
             & 0.95 & -46.7 &0.64 & - & - & - \\
\hline
$\Phi^{*}_s$& - & - & -  & 2.95 & -4.7 & 1.33 \\
             & - & - & - & 3.05 & -14.5 & 0.78 \\
             & - & - & - & 3.1 & -21.4 & 0.66\\
             & - & - & - & 3.15 & -29.6 &0.57 \\
\hline
$\Phi^{*}_{s1}$ & 1.8 & -6.1 & 1.29 & - & - & - \\
                & 1.9& -14.6 &0.88 &-&-&- \\
                & 2.0 & -26.0 &0.69 &-&-&- \\
                & 2.1 &-40.0 &0.58 &-&-&- \\
\hline\hline
\end{tabular}
\\
$g'=1.32,\,\beta'=1.12,\,\lambda'=0.56,\,g'_\sigma=0.761$ \\
\begin{tabular}{c|ccc|ccc}
\hline\hline
&& $c=+1$ &&& $c=-1$ \\
\hline
states & $\Lambda$ & $E$ & $r_{rms}$ & $\Lambda$ & $E$ & $r_{rms}$ \\
\hline
$\Phi^{*}$ & - & -& -& 1.5 & -10.4 & 1.03 \\
           & - & -& -& 1.6 & -18.7 &0.81  \\
           & - & -& -& 1.7 & -30.0 &0.66 \\
\hline
$\Phi^{*}_8$ & 0.8 & -55.0 & 0.63 & - & - & - \\
             & 0.825 & -67.5 & 0.58 & - & - & - \\
             & 0.85 & -82.1 & 0.54 & - & - & - \\
             & 0.875 & -98.8 &0.51 & - & - & - \\
\hline
$\Phi^{*}_s$ & - & - & - & 2.25 & -7.9 & 1.05 \\
             & - & - & -& 2.3 & -14.5 & 0.80  \\
              & - & - & -& 2.35 & -23.0 & 0.65 \\
              & - & - & -& 2.4 & -33.5 &0.55 \\
\hline
$\Phi^{*}_{s1}$ & 1.45 & -5.1 & 1.37 & - & - & - \\
                & 1.5& -11.9 &0.95 &-&-&- \\
                & 1.55 & -21.1 &0.75 &-&-&- \\
                & 1.6&-32.5 &0.63 &-&-&- \\
\hline\hline
\end{tabular}

\end{table}

\newpage

\section*{Acknowledgment}
The authors thank Wei-Zhen Deng for useful
discussions. This project is supported by the National Natural
Science Foundation of China under Grants No. 10625521, No.
10721063, No. 10705001, the Ministry of Science and Technology of China
(2009CB825200) and the Ministry of Education of China (FANEDD under Grants No. 200924, DPFIHE under Grants No. 20090211120029, NCET under Grants No. NCET-10-0442).


\newpage
\appendix
\section{Possible molecular states composed of a pair of excited bottom-strange mesons}

We collect the numerical results for the bottomonium-like system
composed of a pair of excited bottom-strange mesons in the
appendix. Since neither $B_{s0}$ nor $B_{s1}$ is observed
experimentally, we follow Ref. \cite{bmass} and use the mass
values $m_{B_{0}}(J^P=0^+)=5.627$ GeV, $m_{B_{1}}(J^P=1^+)=5.674$ GeV, $m_{B_{s0}}(J^P=0^+)=5.718$ GeV and $m_{B_{s1}}(J^P=1^+)=5.765$ GeV.

\begin{table}[htb]
\centering \caption{The variation of the binding energy $E$ (in unit
of MeV) and the root-mean-square radius $r_{rms}$ (in unit of fm) with
the cutoff and the coupling constants for the $B^*_0-\bar{B}^*_0$
system.}
$\beta'=0.84,\,g'_\sigma=0.761$ \\
\begin{tabular}{c|c|c|c}
\hline\hline
states & $\Lambda$ & $E$ & $r_{rms}$ \\
\hline
$\Phi$ & - & - & - \\
\hline
$\Phi_8$ & 1.15 & -3.0 & 1.41 \\
& 1.2 & -6.5 & 1.04 \\
& 1.25 & -11.1 & 0.85 \\
& 1.35 & -23.4 & 0.66 \\
\hline
$\Phi_{s1}$
& 2.2 & -5.4 & 1.00 \\
& 2.3 & -8.3 & 0.84 \\
& 2.4 & -11.6 & 0.73 \\
\hline\hline
\end{tabular}
\\
$\beta'=0.98,\,g'_\sigma=0.761$ \\
\begin{tabular}{c|c|c|c}
\hline\hline
states & $\Lambda$ & $E$ \\
\hline
$\Phi^{\pm,0}$ & - & - & - \\
\hline
$\Phi^0_8$ & 1.11 & -6.0 & 1.08 \\
& 1.15 & -10.8 & 0.87 \\
& 1.19 & -16.8 & 0.75 \\
& 1.23 & -23.9 & 0.66 \\
\hline
$\Phi^0_{s1}$
& 1.9 & -5.9 & 0.98 \\
& 2.0 & -10.5 & 0.78 \\
& 2.1 & -16.2 & 0.66 \\
& 2.2 & -22.8 & 0.58 \\
\hline\hline
\end{tabular}
\\
$\beta'=1.12,\,g'_\sigma=0.761$ \\
\begin{tabular}{c|c|c|c}
\hline\hline
states & $\Lambda$ & $E$ & $r_{rms}$ \\
\hline
$\Phi^{\pm,0}$ & - & - & - \\
\hline
$\Phi^0_8$ & 1.05 & -5.6 & 1.12 \\
& 1.1 & -13.3 & 0.82 \\
& 1.15 & -24.0 & 0.67 \\
& 1.2 & -37.3 & 0.58 \\
\hline
$\Phi^0_{s1}$ & 1.7 & -5.0 & 1.06 \\
& 1.75 & -7.9 & 0.88 \\
& 1.8 & -11.3 & 0.77 \\
& 1.85 & -15.2 & 0.69 \\
\hline\hline
\end{tabular}

\end{table}

\begin{table}[htb]
\centering \caption{The variation of the binding energy $E$ (in
unit of MeV) and the root-mean-square radius $r_{rms}$ (in unit of
fm) with the cutoff and the coupling constant for the
$B_1-\bar{B}_1$ system when only the pseudoscalar meson exchange
is considered. The $B_1-\bar{B}_1$
system is easier to form a bound state than the $D_1-\bar{D}_1$
case. In this table, we only give the result for the
$B_1-\bar{B}_1$ system with the typical coupling constant
$g^\prime=0.80,\,1.06$. We scan the cutoff range $\Lambda\leq 3.1$
GeV.\label{b1b1}}

$g'=0.80,\,\beta'=0,\,\lambda'=0,\,g'_\sigma=0$ \\
\begin{tabular}{c|ccc|ccc|ccccc}
\hline\hline
      & &$J^P=0^+$ && &$J^P=1^+$& && $J^P=2^+$&\\ \cline{2-10}
state & $\Lambda$ & $E$ & $r_{rms}$& $\Lambda$ & $E$ & $r_{rms}$& $\Lambda$ & $E$ & $r_{rms}$\\\hline
$\Phi^{**}$& 1.0 & -6.4 & 0.94 &1.6 &-1.7 &1.52 &- &- &-\\
           & 1.1 & -12.2& 0.73 &1.8 &-6.8 &0.84 &- &- &-\\
           & 1.2 & -20.3& 0.60 &2.0 &-16.5 &0.58 &- &- &-\\
           & 1.3 & -31.2& 0.50 &2.2 &-32.2 &0.44 &- &- &-\\\hline
$\Phi_8^{**}$&-&-&-&-&-&-&0.8&-13.9 &0.74\\
             &-&-&-&-&-&-&0.85&-20.8&0.64\\
             &-&-&-&-&-&-&0.9&-29.7&0.56\\\hline
$\Phi_s^{**}$& 1.7 & -6.2 & 0.81 &2.45 &-3.4 &1.03 &- &- &-\\
             & 1.8 & -17.4 &  0.53 &2.55 &-9.3 &0.65 &- &- &-\\
             & 1.9 & -34.5 &  0.40 &2.65 &-18.2 &0.49 &- &- &-\\\hline
$\Phi_{s1}^{**}$&-&-&-&-&-&-&1.65&-3.0 &1.10\\
                &-&-&-&-&-&-&1.7&-6.7 &0.78\\
                &-&-&-&-&-&-&1.8&-18.3&0.51\\\hline\hline

\end{tabular}
\\
{$g'=1.06,\,\beta'=0,\,\lambda'=0,\,g'_\sigma=0$} \\

\begin{tabular}{c|ccc|ccc|ccccc}
\hline\hline
      & &$J^P=0^+$ && &$J^P=1^+$& && $J^P=2^+$&\\ \cline{2-10}
state & $\Lambda$ & $E$ & $r_{rms}$& $\Lambda$ & $E$ & $r_{rms}$& $\Lambda$ & $E$ & $r_{rms}$\\\hline
$\Phi^{**}$& 0.8 & -18.2 & 0.69 & 1.1 &-6.1 &0.95 &- &- &-\\
           & 0.85 & -24.8&  0.61 & 1.2 &-11.2 &0.74 &- &- &-\\
           & 0.9 & -32.8& 0.55 & 1.3 &-18.4 &0.61 &- &- &-\\\hline
$\Phi_8^{**}$&-&-&-&-&-&-&0.8&-70.2&0.44\\
             &-&-&-&-&-&-&0.825&-81.8&0.41\\
             &-&-&-&-&-&-&0.85&-94.7&0.39\\\hline
$\Phi_s^{**}$& 1.3 & -3.8 & 1.02 &1.8 &-5.0&0.88 &- &- &-\\
             & 1.35 & -9.5 & 0.70 &1.9 &-14.6 &0.56 &- &- &-\\
             & 1.4 & -17.6 &  0.55 &2.0 &-29.3 &0.42 &- &- &-\\\hline
$\Phi_{s1}^{**}$&-&-&-&-&-&-&1.3&-4.1 &0.98\\
                &-&-&-&-&-&-&1.35&-10.0&0.68\\
                &-&-&-&-&-&-&1.4&-18.3&0.54\\\hline\hline

\end{tabular}

\end{table}

\begin{table}[htb]
\centering \caption{The variation of the binding energy $E$ (in unit
of MeV) and the root-mean-square radius $r_{rms}$ (in unit of fm) with
the cutoff and the coupling constant for the $B^*_0-\bar{B}_1$
system when only the pseudoscalar meson exchange is
considered.}

$g'=0.80,\,\beta'=0,\,\lambda'=0,\,g'_\sigma=0$ \\
\begin{tabular}{c|ccc|ccc}
\hline\hline
&& $c=+1$ &&& $c=-1$ \\
\hline
states & $\Lambda$ & $E$ & $r_{rms}$ & $\Lambda$ & $E$ & $r_{rms}$ \\
\hline
$\Phi^{*}$ & - & - & - & 1.6 & -1.8 & 1.50 \\
           & - & - & - & 1.8 & -6.9 & 0.84 \\
             & - & - & - & 2.0 & -16.5 & 0.58 \\
             & - & - & - & 2.2 & -32.2 & 0.44 \\
\hline
$\Phi^{*}_8$ & 0.8 & -14.3 & 0.74 & - & - & - \\
             & 0.825 & -17.6 & 0.68 & - & - & - \\
             & 0.875 & -25.6 & 0.59 & - & - & - \\
             & 0.9 & -30.3 & 0.56 & - & - & - \\
\hline
$\Phi^{*}_{s}$ & - & - & - & 2.45 & -3.5 & 1.02 \\
           & - & - & - & 2.55 & -9.5 & 0.65 \\
             & - & - & - & 2.65 & -18.4 & 0.49 \\
             & - & - & - & 2.75 & -30.4 & 0.40 \\
\hline
$\Phi^{*}_{s1}$ & 1.65 & -3.0  & 1.12& - & - & - \\
                & 1.75 & -11.7 & 0.62 & - & -& -\\
                & 1.8 & -18.1 & 0.52 & - & - & - \\
                & 1.85 & -26.0 & 0.45 & - & - & - \\
\hline\hline
\end{tabular}
\\
$g'=1.06,\beta'=0,\,\lambda'=0,\,g'_\sigma=0$ \\
\begin{tabular}{c|ccc|ccc}
\hline\hline
&& $c=+1$ &&& $c=-1$ \\
\hline
states & $\Lambda$ & $E$ & $r_{rms}$ & $\Lambda$ & $E$ & $r_{rms}$ \\
\hline
$\Phi^{*}$ & - & - & - & 1.05 & -4.4 & 1.09 \\
           & - & - & - & 1.25 & -14.8 & 0.66 \\
             & - & - & - & 1.35 & -23.2 & 0.56 \\
             & - & - & - & 1.45 & -34.2 & 0.48 \\
\hline
$\Phi^{*}_8$ & 0.8 & -71.4 & 0.44 & - & - & - \\
             & 0.825 & -83.0 & 0.41 & - & - & - \\
             & 0.85 & -95.9 & 0.39 & - & - & - \\
             & 0.875 & -110.1 & 0.37 & - & - & - \\
\hline
$\Phi^{*}_{s}$ & - & - & - & 1.75 & -2.3 & 1.26 \\
           & - & - & - & 1.85 & -9.6 & 0.67 \\
             & - & - & - & 1.95 & -21.8 & 0.48 \\
             & - & - & - & 2.05 & -39.3 & 0.38 \\
\hline
$\Phi^{*}_{s1}$ & 1.3 & -4.1  & 0.98 & - & - & - \\
                & 1.35 & -9.9 & 0.68 & - & -& -\\
                & 1.4 & -18.2 & 0.54 & - & - & - \\
                & 1.45 & -29.1 & 0.45 & - & - & - \\
\hline\hline
\end{tabular}
\\
$g'=1.32,\,\beta'=0,\,\lambda'=0,\,g'_\sigma=0$ \\
\begin{tabular}{c|ccc|ccc}
\hline\hline
&& $c=+1$ &&& $c=-1$ \\
\hline
states & $\Lambda$ & $E$ & $r_{rms}$ & $\Lambda$ & $E$ & $r_{rms}$ \\
\hline
$\Phi^{*}$ & - & - & - & 0.8 & -6.9 & 0.96 \\
           & - & - & - & 0.9 & -14.2 & 0.73 \\
             & - & - & - & 1.0 & -24.9 & 0.58 \\
             & - & - & - & 1.1 & -39.5 & 0.49 \\
\hline
$\Phi^{*}_8$ & 0.8 & -177.2 & 0.33 & - & - & - \\
             & 0.81 & -187.1 & 0.32 & - & - & - \\
             & 0.82 & -197.3 & 0.31 & - & - & - \\
             & 0.83 & -208.1 & 0.31 & - & - & - \\
\hline
$\Phi^{*}_{s}$ & - & - & - & 1.4 & -4.6 & 0.94 \\
           & - & - & - & 1.5 & -9.7 & 0.68 \\
             & - & - & - & 1.6 & -25.8 & 0.46 \\
             & - & - & - & 1.65 & -36.8 & 0.40 \\
\hline
$\Phi^{*}_{s1}$ & 1.1 & -3.1  & 1.13 & - & - & - \\
                & 1.15 & -10.0 & 0.70 & - & -& -\\
                & 1.2 & -20.9 & 0.53 & - & - & - \\
                & 1.25 & -35.8 & 0.43 & - & - & - \\
\hline\hline
\end{tabular}

\end{table}


\begin{thebibliography}{99}
\bibitem{2003-Choi-p262001-262001}
  S.~K.~Choi et al.,
     Phys.\ Rev.\  Lett {\bf 91}, 262001 (2003).

\bibitem{2005-Choi-p182002-182002}
 S.~K.~Choi et al.,
     Phys.\ Rev.\  Lett {\bf 94}, 182002 (2005).


\bibitem{2005-Aubert-p142001-142001}
 B.~Aubert et al., BABAR Collaboration,
     Phys.\ Rev.\  Lett {\bf 95}, 142001 (2005).


\bibitem{2007-Yuan-p182004-182004}
 C.~Z.~Yuan et al., Belle Collaboration,
     Phys.\ Rev.\  Lett {\bf 99}, 182004 (2007).


\bibitem{2006-Uehara-p82003-82003}
 S.~Uehara et al.,
     Phys.\ Rev.\  Lett {\bf 96}, 082003 (2006).

\bibitem{2007-Abe-p82001-82001}
 K.~Abe et al.,
     Phys.\ Rev.\  Lett {\bf 98}, 082001 (2007).

\bibitem{2007-Aubert-p212001-212001}
 B.~Aubert et al., BABAR Collaboration,
     Phys.\ Rev.\  Lett {\bf 98}, 212001 (2007).

\bibitem{2007-Wang-p142002-142002}
 X.~L.~Wang et al., Belle Collaboration,
     Phys.\ Rev.\  Lett {\bf 99}, 142002 (2007).

\bibitem{2008-Choi-p142001-142001}
 S.~K.~Choi et al., Belle Collaboration,
     Phys.\ Rev.\  Lett {\bf 100}, 142001 (2008).

\bibitem{2008-Mizuk-p72004-72004}
 R.~Mizuk et al., Belle Collaboration,
     Phys.\ Rev.\  D {\bf 78}, 072004 (2008).

\bibitem{2009-Aaltonen-p242002-242002}
 T.~Aaltonen et al., CDF Collaboration,
     Phys.\ Rev.\  Lett {\bf 102}, 242002 (2009).

\bibitem{2004-Swanson-p197-202}
 E.~S.~Swanson, Phys.\ Lett. B {\bf 598}, 197 (2004);~ E.~S.~Swanson, Phys.\ Lett. B {\bf 588}, 189 (2004);
 ~T.~Fernandez-Carames, A.~Valcarce, and J.~Vijande,  Phys.\ Rev.\  Lett {\bf 103}, 222001 (2009);
 F. Close and C. Downum, Phys. Rev. Lett. 102, 242003 (2009); F. Close, C.
 Downum, and C. E. Thomas, arXiv:1001.2553v1 [hep-ph].

\bibitem{2008-Liu-p94015-94015}
 Su Houng Lee, ~A.~Mihara,~F.~ Navarra, and M.~ Nielsen,  Phys. Lett. B {\bf{661}}, 28 (2008);
 ~C.~Meng and K.~T.~Chao,~arXiv:0708.4222[hep-ph];~G.~J.~Ding,~arXiv:0711.1485[hep-ph].

\bibitem{2009-Liu-p17502-17502}
 N.~Mahajan  Phys.\ Lett.\  B {\bf 679}, 228 (2009);
 ~T.~Branz,~T.~Gutsche, and V.~E.~Lyubovitskij, Phys.\ Rev.\  D {\bf 80}, 054019 (2009);
 ~G.~J.~Ding, Eur.\ Phys.\ J.\ C {\bf 64} 297 (2009).

\bibitem{liu1}X. Liu, Z. G. Luo, Y. R. Liu, Shi-Lin Zhu, Eur.
Phys. J. C 61, 411 (2009), arXiv:0808.0073 [hep-ph].

\bibitem{liu2}Y. R. Liu, X. Liu, W. Z. Deng, Shi-Lin Zhu, Eur.
Phys. J. C 56, 63 (2008), arXiv:0801.3540 [hep-ph].

\bibitem{liu3}X.~Liu,~Y.R.~Liu,~W.~Z~Deng, and S.~L. Zhu,  Phys.\ Rev.\  D {\bf 77}, 094015
(2008).

\bibitem{liu4}X.~Liu,~Y.R.~Liu,~W.~Z~Deng, and S.~L. Zhu,  Phys.\ Rev.\  D {\bf 77}, 034003
(2008).

\bibitem{liu5} X.~Liu and S.~L.Zhu,  Phys.\ Rev.\  D {\bf 80}, 017502
(2009).

\bibitem{pdg}
 C.~Amsler et al., (Particle Data Group), Phys.\ Lett. B {\bf 667}, 1 (2008).

\bibitem{hpz}P. Z. Huang, L. Zhang, Shi-Lin Zhu, Phys. Rev. D 80,
014023 (2009).

\bibitem{bmass}W. A. Bardeen, E. J. Eichten, C. T. Hill, Phys. Rev. D 68, 054024
(2003); P. Colangelo, F. De Fazio, R. Ferrandes, Nucl. Phys. Proc.
Suppl. 163, 177 (2007).

\end{thebibliography}
\end{document}